\documentclass[aps,prd,10pt,amsmath,floats,floatfix, twocolumn,superscriptaddress,nofootinbib,showpacs, reprint,longbibliography]{revtex4-2}
\usepackage{graphicx} 
\usepackage{mathrsfs}
\usepackage{amsmath, amssymb, graphicx}
\usepackage[pagebackref=false, colorlinks=true]{hyperref}
\hypersetup{colorlinks,linkcolor=blue,urlcolor=blue}

\newcommand{\p}{\partial}

\begin{document}
\preprint{}

\title{Homothetic Killing horizons in generic Vaidya spacetimes}

\author{Ritwika Ghoshal}
\email{ritwikaghoshal1997@gmail.com}
\affiliation{Department of Physics, Indian Institute of Technology Kanpur, Kalyanpur, Kanpur-208016, Uttar Pradesh, India.}

\author{Nilay Kundu}
\email{nilayhep@iitk.ac.in}
\affiliation{Department of Physics, Indian Institute of Technology Kanpur, Kalyanpur, Kanpur-208016, Uttar Pradesh, India.}

\author{Srijit Bhattacharjee}
\email{srijuster@gmail.com}
\affiliation{Indian Institute of Information Technology Allahabad (IIITA), Devghat, Jhalwa, Prayagraj-211015, Uttar Pradesh, India.}

\date{\today}

\begin{abstract}
We study the conformal Killing equation for generic Vaidya-like spacetimes, including those with rotation. We show that these spacetimes admit a unique class of conformal Killing vectors that are homothetic for mass, charge, or rotation parameters being linear functions of the advanced null-time. For the Kerr-Vaidya metric, the solution to the conformal Killing equation exists iff both mass and rotation parameters become dynamic.  The presence of a homothetic Killing vector (HKV) for such a spacetime enables one to conformally map the original dynamical spacetime to a stationary spacetime, enabling access to the standard methods pertaining to a Killing horizon. The surface where an HKV becomes null is termed the homothetic Killing horizon. We discuss the thermodynamic properties of such homothetic Killing horizons and formulate a version of the first law (or flux balance law) for spherically symmetric Vaidya spacetimes. We further study the maximal analytic extension of a charged Vaidya metric and indicate its implications for studying particle creation in such backgrounds.  

\end{abstract}

\maketitle
\section{Introduction}
Black Holes are considered to be the most intriguing objects in the Cosmos, offering tons of interesting phenomena that baffled scientists for decades. On the other hand, a large amount of information regarding the spacetime structure, including the gravitational waves are uncovered when processes involving black holes are studied \cite{PhysRevLett.116.061102, PhysRevLett.125.101102, Akiyama_2019}. In its simplest form, when a black hole is considered to be almost stagnant, it is described by only a few parameters, namely mass, charge, and angular momentum. These parameters organize themselves mysteriously to reveal a thermodynamic behavior. The laws of black hole mechanics have gained tremendous attention after Bekenstein and Hawking proposed that stationary black holes have entropy \cite{bardeen1973four, bekenstein1973black, hawking1975particle}. Many believe that the root of the quantum nature of gravity is hidden in the understanding of this thermodynamic behavior of such a geometric entity, admitting the event horizon as a one-way membrane separating the world from a causally disconnected region. The intrinsic features of the event horizon of a stationary black hole are well studied \cite{Wald:1999vt}. However, black holes in nature are usually not inert, but they should become dynamic. The structure of such dynamical black holes has many similarities with the stationary black holes, but some crucial differences do exist. One of the major identifications for such a configuration is the absence of a global time-like Killing vector (KV). KVs played a crucial role in the development of black hole thermodynamics. In the static scenario, such a Killing vector not only specifies a conserved quantity, but also provides a notion of a {\it local} horizon where its norm vanishes, known as a Killing horizon (KH). Unlike the event horizon, which is a global concept and possesses a teleological nature, KH is free from such issues. For any stationary black hole, the event horizon becomes a Killing horizon \cite{Hawkingellisbook}. Spacetimes admitting a Killing vector are useful to study the properties of solutions to Einstein's equations \cite{Heusler}. They are also a very crucial ingredient to study black hole perturbations \cite{Pound:2021qin}. 

In non-stationary spacetimes, a KH ceases to exist, and one loses many of the advantages a KV provides. Non-stationary spacetimes are inevitable in the description of black hole formation due to gravitational collapse \cite{PhysRev.56.455}. How to satisfactorily define temperature and establish the thermodynamical laws for such dynamical black holes has still not been satisfactorily understood. In this connection, isolated and dynamical horizons \cite{Ashtekar:2004cn}, trapping horizons \cite{PhysRevD.49.6467}, and slowly-evolving horizons \cite{PhysRevLett.92.011102} have been proposed and received quite a bit of success in understanding some of the issues outlined above. However, dynamical or trapping horizons may not always have the null character. This introduces subtleties in establishing thermodynamical relations on such horizons. Further, beyond slow evolution, when a black hole is far from equilibrium, many of these setups fail to work. 
 
A very useful model for studying nonstationary black holes is the Vaidya solution \cite{Vaidya0, vaidya1}, which describes a black hole (BH) formed by a collapsing (or radiating) null dust. Vaidya solutions are used to study a wide range of topics, starting from the gravitational collapse and singularities of BHs to the Hawking radiation \cite{IHDwibedi1, PhysRevD.23.2813}.  Vaidya spacetimes do not admit a KH. However, as we show in this article, a generalization of the KH, namely the conformal Killing horizon (CKH), can be obtained for Vaidya spacetimes under certain specific conditions. Conformal Killing vectors are solutions of the conformal Killing equation- $\mathcal{L}_\xi g=\Omega g$. Like KH, a CKH is also a null surface where the norm of a conformal Killing vector (CKV) vanishes \cite{dyer1979conformal, Jacobson:1993pf, sultana2004conformal}.  The concept of CKH was introduced to explore whether results obtained for stationary BHs could be generalized to non-stationary cases. It is known that for a linear mass function, the Schwarzschild-Vaidya solution admits a CKH \cite{galaxies2010062, Nielsen:2017hxt}. The charged version (Vaidya-Bonnor metric \cite{Bonnor:1970zz}) also shares a similar feature \cite{Tarafdar:2022rzz, Koh:2020hta}.  The Kerr–Vaidya metrics are the simplest extensions that allow for an axially symmetric and time-dependent mass distribution \cite{Vaidya:1973zza}. Since real astrophysical objects rotate, it is also important to check whether there exists CKH for the rotating case, even in the spirit of a toy model.

In this note, we have shown that a wide class of Vaidya metrics admits a special class of CKVs- namely the Homothetic Killing Vectors (HKVs). HKVs are solutions of the conformal Killing equation with a constant conformal factor. We establish a wide class of Vaidya metrics, such as the Husain type spacetimes (a class of generalized Vaidya solution) \cite{Husain:1995bf}, and the Kerr-Vaidya spacetimes admit a unique kind of HKV.  Further, these metrics admit solutions for HKVs only when the mass, charge, or rotation parameter (angular momentum per unit mass) depends linearly on the null coordinate $ v$. We have also found that there is no solution for CKV if we don't turn on the $v$ dependence of both the mass and rotation parameters. In these spacetimes, the HKVs also provide a corresponding version of null boundary - they are termed as Homothetic Killing Horizons (HKHs). Due to the existence of an HKH, one can map these dynamical spacetimes to a static or stationary one, where an HKV becomes a KV. We compute the surface gravity for the HKHs and comment on the relation between them and the thermodynamic temperatures. The thermodynamic laws for HKHs are discussed. We also provide the necessary setup for studying particle creation through the Hawking effect near HKHs. 

The paper is organized as follows. In \S\,\ref{sec_basic set up}, we describe the conventions that will be used and define several basic elements, such as the CKV, horizons, and the associated notions of surface gravity with it. Next, in \S\,\ref{sec_CKH_spherical_Vaidya} we analyze the existence of HKVs for generic spherically symmetric Vaidya space times, including Husain-type BHs. Following that, in \S\,\ref{sec_Kerr_Vaidya}, we study the HKVs admitted by rotating versions of Vaidya BHs. In \S\,\ref{sec_thermo}, we discuss the thermodynamic properties of the CKH in Kerr-Vaidya BHs and develop a description of a self-similar Reissner-Nordström-Vaidya spacetime using the setup of CKH. Finally, we conclude in \S\,\ref{sec_conclusions}. 

\section{Notations \& Basic Set Up} \label{sec_basic set up}

Throughout the paper, we use spacetime metrics of signature $(D-1); (-1,1,1,1,\cdots)$ for $ D$-dimensional spacetimes. Unless otherwise specified, we will focus on $4$ dimensions throughout this note. The spacetime indices are denoted by Latin letters. We adopt $G=c=1.$

\subsection{Conformal Killing horizon} \label{basic_set_up_CKH}
For a stationary spacetime metric $g_{ab}$, admitting a Killing vector field $\xi^a$, the flow generated by $\xi^a$ leaves the metric invariant. The Killing equation reads,
\begin{equation}\label{KE}
    \mathcal{L_\xi} g_{ab}= \nabla_a \xi_b +\nabla_b \xi_a = 0 \, , 
\end{equation}
where $\xi_a= g_{ab} \xi^b$ and $\nabla_a$ is the covariant derivative operator associated with $g_{ab}$.  The KH of $g_{ab}$ is a null-surface upon which the norm of $\xi^a$ vanishes, i.e. $\xi^a \xi_a|_{KH} = 0$. 

A CKV $\xi$, on the other hand, leaves the metric $g_{ab}$ invariant up to a conformal factor, 
\begin{equation}\label{CKE1}
    \mathcal{L}_\xi g_{ab}= 2\lambda g_{ab}=\frac{2}{D}(\nabla \cdot \xi)g_{ab} \, ,
\end{equation}
such that $\lambda$ is some function of the space-time coordinates. 

Now, consider a metric $\bar{g}_{ab}$ that is related to $g_{ab}$ via a conformal transformation, 
\begin{equation}\label{cnftr}
  \bar{g}_{ab}= \Omega^2 g_{ab}, \,\,\,\,\,\,
  \bar{g}^{ab}= \Omega^{-2} g^{ab}.
\end{equation}
In this case, assuming $\xi^a$ is a Killing vector for $g_{ab}$, i.e., $\mathcal{L}_\xi g_{ab}=0$, 
\begin{equation}
    \mathcal{L}_\xi \bar{g}_{ab}= \mathcal{L}_\xi (\Omega^2 ) g_{ab} = \mathcal{L}_\xi\left( \text{ln}\, \Omega^2 \right)\bar{g}_{ab}.
\end{equation}
If $\Omega^2$ is not constant along the Killing vector's orbit, $\xi^a$ will be CKV for $\bar{g}_{ab}$, satisfying the conformal Killing equation (CKE)
\begin{equation}\label{CKE2}
     \mathcal{L_\xi} \bar{g}_{ab}= \xi^c \bar{\nabla}_c\left(\text{ln}\, \Omega^2 \right)= \bar{\nabla}_a \bar{\xi}_b +\bar{\nabla}_b \bar{\xi}_a  = \frac{2(\bar{\nabla}_c \xi^c)}{D} \bar{g}_{ab} \, ,
\end{equation}
where, $\bar{\nabla}$ is the covariant derivative associated with $\bar{g}_{ab}$, and
\begin{equation}
    \bar{\xi}_a =\bar{g}_{ab}\xi^a = \Omega^2 \xi_a, \quad \quad \bar{\xi}^a =\bar{g}^{ab}\bar{\xi_a} =\xi^a.
\end{equation}
Thus, we learn that if $\xi^a$ is a Killing vector for $g_{ab}$, the same $\xi^a$ will be a CKV for the conformally related metric $\bar{g}_{ab}$. On the contrary, given that $\xi^a$ is a CKV for $\bar{g}_{ab}$, i.e., $\mathcal{L}_\xi \bar{g}_{ab} = 2 \lambda \bar{g}_{ab}$, one can check that (see \cite{Jacobson:1993pf}) if 

\begin{equation}\label{conf-factor}
    \mathcal{L}_\xi (\Omega^2) - 2 \lambda \Omega^2 = 0, 
\end{equation} 

 then $\xi^a$ will be a Killing vector for $g_{ab}$, which is conformally related to $\bar{g}_{ab}$ as in equation \eqref{cnftr}. 

A CKH is defined as the null hypersurface where the CKV becomes null with a vanishing norm. Therefore, the metric $\bar{g}_{ab}$ will admit a CKH when 
\begin{equation} \label{ckh_condition}
\bar {\xi}^a \bar{\xi}_a|_{CKH}=  \xi^a \bar{\xi}_a|_{CKH} = 0. 
\end{equation}

\subsection{Conformally invariant surface gravity} \label{IIB}
 
Under a conformal transformation, equation \eqref{cnftr}, the Einstein equation does not remain invariant. Therefore, it is expected that the properties of the KH of a BH solution of Einstein gravity may not automatically be carried forward. However, the causal structure remains invariant under conformal transformation. Motivated by this, a definition of the conformally invariant temperature of BHs that admits a CKH can be constructed \cite{Jacobson:1993pf, sultana2004conformal}. 

To define a notion of surface gravity for conformal KHs is subtle. Recall that for a static or stationary spacetime comprising a true KH generated by the Killing vector $\xi^a$, the following definitions of the surface gravity of that KH are all equivalent 
\begin{equation}
\begin{split}
   & \nabla_a \left(\xi^b\xi_b\right)= -2 \kappa_1 \xi_a \, , \quad \xi^b\nabla_b\xi^a = \kappa_2 \xi^a \, , \\
    & -\frac{1}{2} \left(\nabla^a \xi^b\right) \left(\nabla_{[a} \xi_{b]}\right) =(\kappa_3)^2 \, .
\end{split}
\end{equation}
However, for a CKH, in cases where $\xi^\alpha$ is a CKV of the spacetime metric $\bar{g}_{ab}$ which is conformally related to $g_{ab}$ by equation \eqref{cnftr}, we can have the following candidates for surface gravity \footnote{Note that in equation \eqref{kappa_1}, \eqref{kappa_2}, \eqref{kappa_3} we have used $\bar{\xi}^a = \xi^a$.}
\begin{subequations}
\begin{align}
    \bar{\nabla}_a \left(\xi^b\bar{\xi}_b\right)= &-2 \,  \bar{\kappa}_1 \, \bar{\xi}_a \, , \label{kappa_1}\\
    \xi^b \bar{\nabla}_b\xi^a = & \, \bar{\kappa}_2 \,  \xi^a \, ,  \label{kappa_2}\\
    -\frac{1}{2} \left(\bar{\nabla}^a \xi^b\right) \left(\bar{\nabla}_{a} \bar{\xi}{_b}\right) =& \, \bar{\kappa}_3^2\,  \label{kappa_3} \, ,
\end{align}
\end{subequations}
which are not equivalent for a CKH, see \cite{Jacobson:1993pf}. 

Furthermore, a little manipulation with equation \eqref{cnftr} to \eqref{ckh_condition} yields the following relations 
\begin{equation}\label{kappa11relation}
    \bar{\kappa}_1=\kappa_1 \, , 
\end{equation}
and
\begin{equation} \label{kappa123relation}
    \bar{\kappa}_1 = \bar{\kappa}_2 - 2\lambda = \bar{\kappa}_3 -\lambda \, .
\end{equation}
Motivated by equation \eqref{kappa11relation}, $\kappa_1 = \bar{\kappa}_1$ can be thought of as the conformally invariant description of surface gravity \cite{Nielsen:2017hxt}. Alternatively, in \cite{Nielsen:2017hxt}, this has been called the geometric surface gravity of the corresponding KH or CKH.
 
In contrast, in \cite{sultana2004conformal}, it was argued that $\bar{\kappa}_2$ should be considered a better candidate to define surface gravity, since it measures the extent to which the parameterization of the null geodesic congruence, represented by the conformal Killing trajectories on CKH, is not affine. Therefore, from equation \eqref{kappa123relation}, we can see that an important difference between the properties of surface gravity on the KHs and that on the CKHs is that in the former case, the surface gravity along each Killing trajectory on the KH is constant, while in the latter case it scales up or down with $\lambda$ along the conformal Killing trajectories generating the CKH. Additionally, in \cite{sultana2004conformal}, a proposal for the zeroth law has been put forward, showing that 
\begin{equation}
    \mathcal{L}_\xi[\bar{\kappa}_2- 2\lambda] = 0.
\end{equation}

\section{Conformal Killing horizon for spherical-Vaidya solutions} \label{sec_CKH_spherical_Vaidya}
We start with a review of the existence of CKH for Vaidya spacetime. 

\subsection{CKH in Schwarzschild-Vaidya spacetime}

When a black hole spacetime becomes dynamical, the idea of an event horizon cannot be given by the KH anymore. There is a certain class of dynamical spacetimes for which the metric admits a conformal transformation that maps the spacetime to a static one. Therefore, the KH of the conformally static spacetime will correspond to the CKH of the dynamical spacetime. Then the Killing vectors in the static spacetime become the conformal Killing vectors of the dynamical spacetime, and the CKH is defined by a null hypersurface where the norm of the CKV vanishes to zero. One of the familiar examples of this kind of dynamical spacetime is the Vaidya spacetime emitting or absorbing null dust.

The conformal Killing horizons and their thermodynamic properties have been described in \cite{Nielsen:2017hxt} for Schwarzschild-Vaidya spacetime accreting matter (null dusts). The relevant spacetime metric is 
\begin{equation} \label{schv}
   ds^2 = -\left[1-\frac{2 m(v)}{r}\right] dv^2 + 2\,dv\,dr + r^2 \left(d\theta^2 + \sin^2 \theta \, d\phi^2 \right) \, . 
\end{equation}
In writing \eqref{schv}, and in the following analysis, we will be denoting the dynamical spacetime with the unbarred metric $g_{ab}$. Also, all our calculations in the subsequent sections will be performed with the dynamical Vaidya metric, not in the conformally transformed static or stationary geometries. However, we used a different convention previously in \S\,\ref{basic_set_up_CKH}, where the metric for the dynamical BH spacetime was written with barred quantities $\bar{g}_{ab}$. The CKE given in equation \eqref{CKE2} is satisfied by the following CKVs
\begin{equation} \label{ckv_schv}
    \xi^a = \left(\frac{m(v)}{M},\frac{\mu r}{M},0,0\right) \, ,
\end{equation}
for the dynamical Vaidya metric written in \eqref{schv}, subjected to the linear mass function 
\begin{equation}
    m(v) = M +\mu (v-v_0).
\end{equation}
The CKV components were chosen such that in the static limit ($\mu=0$), $\xi$ reduces to a Killing vector field of Schwarzschild spacetime with unit norm at asymptotic infinity.
Accordingly, using equation (\ref{ckh_condition}), the CKH was found to be located at 
\begin{equation}
     r_{CKH}= \frac{m(v)}{4 \mu}\left[1\pm \sqrt{1-16\mu}\right]\, .
 \end{equation}
This dynamical spacetime can be mapped to a static one using a conformal transformation, and, following equation \eqref{cnftr}, the corresponding conformal factor $\Omega^2$ can be chosen as 
 \begin{equation}
     \Omega^2 = \frac{m(v) \, r}{2M^2}.
 \end{equation}
 This choice of the conformal factor is not unique; it has a normalization freedom under rescaling by a constant. In conformally static spacetime, the vector in equation \eqref{ckv_schv} will be a timelike Killing vector satisfying equation \eqref{KE}. The relevant metric is,
 \begin{equation}\label{conf-trans}
 \begin{split}
     & \Omega^{-2} \, g_{ab} \, dx^a dx^b =\frac{2M^2}{m(v)r} \bigg[2 \, dv \, dr -   \\  & \quad  \left(1-\frac{2 m(v)}{r}\right) dv^2 \quad + r^2 \left(d\theta^2 + \sin^2\theta d\phi^2 \vphantom{\frac{2 m(v)}{r}} \right) \bigg] \, .
 \end{split}
 \end{equation}
 Note that the metric in \eqref{conf-trans} is obtained by doing the conformal transformation starting from the dynamical metric \eqref{schv}, 
 \begin{equation} \label{conf_dynamic_to_static}
g_{ab} \rightarrow \Omega^{-2} g_{ab} \, \, ,
 \end{equation}
which, as we declared earlier, follows a different convention than \eqref{cnftr}. Also, the spacetime in \eqref{conf-trans} possesses a Killing vector, but is not asymptotically flat, since $\Omega$ does not go to unity at infinity. However, focusing on the metric in \eqref{schv}, for the mass function
 \begin{equation} \label{mf2}
     m(v)= \mu v \, ,
 \end{equation}
the CKV is $\xi^a = (v,r,0,0)$, and the location of the CKH turns out to be
 \begin{equation} \label{kh2}
     r \big|_{CKH}= \frac{v}{4}\left[1\pm \sqrt{1-16\mu}\right] \, ,
 \end{equation}
which is real given 
 \begin{equation*}
 \mu \leq \frac{1}{16} \, .
 \end{equation*}
Hence, we see that a conformal map relates a dynamical spacetime (\ref{schv}) to a static spacetime \eqref{conf-trans}. One can use this HKV or the Killing vector to analyze the geodesic structure, shadows, temperature, etc., for such spacetimes \cite{Heydarzade:2023gmd, Vertogradov:2022yja}.  \\

\subsection{HKV in spherical-Vaidya spacetimes: Kinematics} \label{HKV_Husain_Kinematics}
  Here, we consider a broader class of Vaidya metrics with spherical symmetry, beyond the Schwarzschild-Vaidya BHs discussed in the previous subsection. We will focus on general models of such a spacetime, which is known as the Husain-type BH geometries \cite{Husain:1995bf}, for which the metric can be written in Eddington-Finkelstein (EF) type coordinates as 
  \begin{align} \label{HVMetric}
     d{s}^2 = &-\left[1-\frac{2 m(v)}{r} + \frac{(\alpha(v))^{2k}}{r^{2k}}\right] dv^2 + 2\,dv\,dr\\ \nonumber &+ r^2 \left(d\theta^2 + \sin^2 \theta ~d\phi^2 \right) \, ,   
\end{align}
where $k$ is a constant parameter. Note that for the well-known Reissner-Nordstrom BH with electro-magnetic charges, the parameter takes the value $k=2$ \footnote{Although in \eqref{HVMetric} we are considering Vaidya versions of the Husain-type metrics, where both mass and the charges are functions of $v$}. Notably, we will get the well-known Reissner-Nordstrom BH with electro-magnetic charges. The Husain-type metrics consider infalling fluid onto the BH.  The fluid equation of state satisfying $P=k\rho,$ where $P$ and $\rho$ are fluid pressure and energy density, respectively. $k> {1\over 2}$ corresponds to the asymptotic flat solutions to which we restrict our focus. For $k=1$ and $\alpha=q$, the metric coincides with the charged Vaidya solution \cite{PhysRevD.95.124033}. The stress-energy tensor for the geometry can be written in terms of the outgoing and ingoing null vectors as:
\begin{equation} \label{Tab_hussain}
    T_{ab}=\mu n_an_b+\rho (l_an_b+l_bn_a)-P(g_{ab}+l_an_b+l_bn_a) \, .
\end{equation}
Here $\mu$ corresponds to the flux of energy flowing inward along the null direction $n^a=\{0,-1,0,0\}.$ The radially outward null vector is denoted by $l^a=\{1,f/2,0,0\},$ and $g_{ab}$ denotes the components of the Husain type metric. The expressions for the fluid stress energy tensor parameters can be cast in the following form: 
\begin{equation}
    \mu=\frac{ \partial_v m}{4\pi r^2} -\frac{\partial_v\psi}{8\pi r^{2k+1}} , \;  \rho = \frac{\psi(2k-1)}{8\pi r^{2k+2}}, \; P=\frac{k(1-2k)\psi}{8\pi r^{2k+2}} ,  \label{EC}
\end{equation}
where $\psi(v)=(\alpha(v))^{2k}.$ The energy conditions on the matter sector restrict the particular forms of $m$ and $\alpha$. To ensure the weak, null, dominant, and strong energy conditions all hold, one must have:
\begin{equation}
    \mu \geq 0 \; , \; \;  \rho \geq 0 \; \mbox{and} \; \; 0 \leq P \leq \rho \, .  \label{EC1}
\end{equation} 

Next, our goal is to investigate the conditions under which the metric in \eqref{HVMetric} will have CKVs that will satisfy the CKE. The CKE, which, following its definition given in \eqref{CKE2}, can be written in our newly adapted convention as 
\begin{equation} \label{CKE3a}
  \mathcal{L_\xi} g_{ab}= \nabla_a \xi_b+\nabla_b \xi_a = \frac{1}{2} (\nabla_c \xi^c) g_{a b}  \, .
\end{equation}
We have checked that the form of $\xi$ written in \eqref{ckv_schv} is a solution to the CKE in \eqref{CKE3a}, given that $m(v)$ and $\alpha(v)$ have a linear dependence on $v$. To be more precise, we found that for 
\begin{equation} \label{malphav1}
    m(v)=\mu \, v,~~\text{and}~~~\alpha(v)=\alpha_1 \, v\, ,
\end{equation}
with $\mu$ and $\alpha_1$ being undertermined constant parameters, the $\xi$ given in \eqref{ckv_schv} satisfies the CKE \eqref{CKE3a}. These parameters depict the rate of change, or accretion, of mass and other `charges'. 

Further, we have also checked that for a different initial condition for the linear $v$-dependence in $m(v)$ and $\alpha(v)$, as written below 
\begin{equation} \label{malphav2}
     m(v) = M +\mu (v-v_0) \, , ~ \text{and} ~ \alpha(v) = \alpha_0 +\alpha_1 (v-v_0) \, ,
\end{equation}
the CKE \eqref{CKE3a} admits a CKV as a solution in the form of $\xi$ given in \eqref{ckv_schv}, provided the following additional constraint for the parameters
\begin{equation} \label{constraint1}
 \alpha_1= \frac{\mu \alpha_0}{M}.   
\end{equation}
get satisfied. 

It is instructive to highlight the significance of the constraint written in \eqref{constraint1}. The $m(v)$ and $\alpha(v)$ profiles written in \eqref{malphav1} and \eqref{malphav2} correspond to two different physical situations in terms of the initial conditions when the $v$-dependence is kicking in. To be more precise, in \eqref{malphav1} we have a Vaidya BH with its dynamics being present eternally. The profile in \eqref{malphav2}, on the other hand, may be associated with a BH that was static initially, but becomes dynamical at a given time $v=v_0$. Thus, the parameters describing the BH were constant, $M$ and $\alpha_0$, until $v=v_0$, and acquired a linear $v$ dependence thereafter, such that $\mu$ and $\alpha_1$ are the rates at which the parameters change with $v$. Now, in such a scenario, we learn that the spacetime will admit a CKV in the form of $\xi$ as in \eqref{ckv_schv}, if the parameters satisfy \eqref{constraint1}. 

After establishing that the Husain type Vaidya black holes in \eqref{HVMetric} admit $\xi$ in \eqref{ckv_schv} as a CKV for either \eqref{malphav1} or \eqref{malphav2} with the condition \eqref{constraint1}, we explore its uniqueness. In other words, we ask the following question, starting with a generic ansatz for the $\xi$ given by  
\begin{equation} \label{CKE_Ansatz1}
     \xi^a =\{\xi^v(v,r),\xi^r(v,r),0,0\} \, ,
\end{equation}
and demanding that it will be a CKV for the metric \eqref{HVMetric}, can we uniquely determine the specific form of $m(v)$, $\alpha(v)$, as given in \eqref{malphav1} and \eqref{malphav2}, and the form of $\xi$ as given in \eqref{ckv_schv}? Most importantly, it will be interesting to see if the constraint in \eqref{constraint1} emerges out of this exercise as a necessary and sufficient condition for the existence of CKV for the profile in \eqref{malphav2}. This approach is more convincing for understanding why the functions $m(v)$ and $\alpha(v)$ take the particular forms mentioned above.

To solve for the CKV, we substitute the ansatz in \eqref{CKE_Ansatz1} into the CKE \eqref{CKE3a} for the Husain metric \eqref{HVMetric}. Following this, the $rr$-component of \eqref{CKE3a} reads $\partial_r \xi^v(v,r) = 0$, which can be solved as 
\begin{equation} \label{HV_CKE_Condition1}
     \xi^v(v,r)=\xi^v(v) + c_1 \, , 
\end{equation}
with $c_1$ being some undetermined constant. From the $vr$, $\theta\theta$ and $\phi\phi$-components of \eqref{CKE3a} we get \begin{equation}
    \partial_r \xi^{r}(v,r)- 2 \xi^r(v,r)/r+ \partial_v \xi(v) =0,
\end{equation} 
which can be solved as 
\begin{equation} \label{eqnxir0}
\xi^r(v,r) = r \left(\partial_v \xi(v)+r c_2(v)\right) \, ,
\end{equation}
such that $c_2(v)$ is another undetermined function. Finally, we substitute them back into \eqref{CKE3a} and examine the remaining $vv$-component of this equation. Additionally, to satisfy this equation, we also need to demand that the coefficients of various powers of $r$ should independently vanish. This results in the condition 
$c_2(v) = 0$ for any non-zero $m(v)$. This in tern gives us $\partial_v^2 \xi(v) =0$, which leads us to 
\begin{equation} \label{xiv_eqn}
\xi^v(v,r) =  c_3+ c_4 \, v \, ,
\end{equation}
establishing that $\xi^v$ must be a linear function of $v$. Using this in \eqref{eqnxir0}, we also obtain that $\xi^r$ is a linear function of $r$
\begin{equation} \label{eqnxir}
\xi^r(v,r) = c_4 \, r \, ,
\end{equation}
Finally, using all the relations so far, we arrive at the important relation involving the functions $m(v)$ and $\alpha(v)$,  
\begin{equation} \label{condition_CKV}
    {\partial_v m(v) \over m(v)} = {\partial_v \alpha(v) \over \alpha(v)} = {\partial_v \xi^v(v) \over \xi^v(v)}\, .
\end{equation}
This condition is important as it puts a constraint on $m(v)$, $\alpha(v)$ in terms of $\xi^v(v)$, and in turn on $\xi^r(r)$ via \eqref{eqnxir}. This should be considered as the most general condition for the Husain metric \eqref{HVMetric} to have a CKV as in \eqref{CKE_Ansatz1}. Also, note that the linear $v$ dependencies of $m(v)$ and $\alpha(v)$, which we discussed previously,  can be understood as a consequence of the condition in \eqref{condition_CKV}, once the linear $v$ dependencies of $\xi^v(v)$ are fixed as in \eqref{xiv_eqn}. Therefore, we see that the choices we made as ansatz previously, as in \eqref{ckv_schv} and \eqref{malphav2}, can be understood as a unique solution coming from the CKE \eqref{CKE3a}. 

Before we proceed, it is worth highlighting that a similar analysis for the existence of CKV was given in \cite{Koh:2020hta} for charged Vaidya spacetime or Reissner-Nordström Vaidya (VRN) spacetime. Note that the latter is a special case of the Husain metric, obtained by putting $k=1$ and $\alpha(v)=q(v)$ (charge function) in (\ref{HVMetric}), and thus the relevant metric for VRN can be written as 
 \begin{equation} \label{metric_VRN}
 \begin{split}
    d{s}^2 = -\left[1-\frac{2 m(v)}{r} + \frac{q^2(v)}{r^2}\right] dv^2 + 2 \, dv \,dr \\  + r^2 \left(d\theta^2 + \sin^2 \theta \, d\phi^2 \right)  \, ,
 \end{split}
 \end{equation}
 where both the mass and the charge are functions of advanced time $v$. The CKVs turned out to be the same as \eqref{ckv_schv}. Both the charge function $q(v)$, and the mass function $m(v)$ were found to be at most linear order in $v$. To be more precise, for the mass and charge functions similar to \eqref{malphav2},
 \begin{equation}
m(v) = M +\mu (v-v_0)\, , ~~ q(v) = q_0 +q_1 (v-v_0) \, ,
 \end{equation} 
 it was shown that $\xi$ as in \eqref{ckv_schv} is a CKV if the following condition holds
 \begin{equation} \label{RNV_condition}
     q_1= \frac{\mu q_0}{M} \, .
 \end{equation}
 The conformal factor, defined in \eqref{cnftr}, was found to be
 \begin{equation}
     \Omega^2=\left(\frac{m(v)}{M} \Tilde{c_4}(v,r)\right)^2,
 \end{equation}
where
 \begin{equation}
    \Tilde{ c_4}(v,r)= \Tilde{c_4}\left(\frac{r}{m(v)}\right),
 \end{equation}
 so that the CKV (\ref{ckv_schv}) satisfies the Killing equation (\ref{KE}) for the conformally static metric. The authors have also computed the surface gravities in this case at CKH, obtained by putting the norm of $\xi^\alpha$ to zero. 
 
\subsection{HKV in spherical-Vaidya spacetimes: Dynamics}\label{HKVform}
Here we provide a proof of the following proposition: If a spherical Vaidya-like spacetime (Husain class) admits an HKV, then it must be of the form $\xi=\{cv, cr, 0, const.\}$. If the spacetime contains a cosmological constant, then $\xi=\{cv+g(r), cr, 0,const,\}$ where $c$ is a constant.\\

We start with the following observations. Recall equation  (\ref{CKE1}),
\begin{equation}\label{CKE3}
     \mathcal{L_{\xi}} g_{ab}= 2\lambda g_{ab}.
\end{equation}
Using this equation, one can show that the following two relations hold for the Ricci tensor and Ricci scalar- 

\begin{equation}\label{EE1}
  \mathcal{L_{\xi}} R_{ab}= -2\nabla_a\nabla_b\lambda - g_{ab}\Box \lambda, \,\,\,\,\,\,
  \mathcal{L_{\xi}} R= -2 \Lambda R -6\Box \lambda.
\end{equation}
Now, the Einstein equation $R_{ab}-\frac{1}{2}g_{ab}R=T_{ab}$ implies,

\begin{equation}\label{EM}
    \mathcal{L_{\xi}} T_{ab}=  -2\nabla_a\nabla_b\lambda  +2g_{ab}\Box \lambda.
\end{equation}

Here, $T_{ab}$ is the energy-momentum tensor.  Also for spacetime admitiing an HKV, 
\begin{equation}\label{TEMeq}
    \mathcal{L_{\xi}} T_{ab}=0.
\end{equation}

 In this case, since $\lambda$ is a constant, from (\ref{CKE3}) we get

\begin{equation}
    \label{div}
    \nabla_a \xi^a=4c,
\end{equation}
where $c=\lambda$ is a constant. Now consider a generic Vaidya spacetime with spherical symmetry. In the EF gauge, the metric would look as,
\begin{equation}\label{sph-metric}
    ds^2=-f(r,v) dv^2+2dv dr+ r^2d\theta^2+r^2\sin^2\theta~ d\phi^2.
\end{equation}

A direct calculation of  $\theta\theta$-component of equation  (\ref{CKE3}) yields:

\begin{align}\label{thetaeq}
   \xi^{a}\partial_a g_{\theta\theta}+2\partial_{\theta}\xi^{\theta}g_{\theta\theta}=2c~ g_{\theta\theta}.
\end{align}

Spherical symmetry implies components of $\xi$ should be independent of $\theta$ and $\phi.$ Further, the homotheticity, together with condition (\ref{div}) rules out any non-zero $\xi^{\theta}$\footnote{Since $\nabla_a\xi^a=\frac{\partial_a(r^2\sin \theta~ \xi^a)}{r^2\sin\theta}.$}. 

Hence, from (\ref{thetaeq}) one can infer

\begin{equation}
    \xi^{r}=c r.
\end{equation}

Next, (\ref{div}) implies
\begin{eqnarray}
     \partial_v\xi^v+2\frac{\xi^r}{r}+ \partial_r\xi^r &=& 4c \nonumber \\ 
   \mbox{or,~~} \partial_v\xi^v &=& c \nonumber\\
    \implies \xi^v &=& c v +g(r) \label{xigen}.
\end{eqnarray}
Note that, up to Eq. \eqref{xigen}, we haven't used any particular form of $f(r,v)$. Hence, the $v$ and $r$ components of $\xi$ are also valid for any $f$ containing a cosmological constant.

Now, using the $vr$-component of equation  (\ref{TEMeq}) we get
\[\partial_r\xi^v T_{vv}=0,\]
which indicates \begin{equation}
    \xi^v=c v.
\end{equation}

One may also add a constant to the form of $\xi^v$. It must also be mentioned that in the presence of a cosmological constant, equation  (\ref{TEMeq}) is no longer valid. Therefore, just assuming a spherical metric for the spatially compact dimensions, the asymptotically flat Vaidya geometries yield a special class of HKVs given by $\xi^a=\{cv+c_0, cr, 0, \chi\}$, where $c_0,\chi$ are constants including zero. 

Now we show that for the metric \eqref{HVMetric}, the only possible $m(v)$ and $\alpha(v)$, for which equation  (\ref{TEMeq}) is satisfied, are linear in $v.$ 
To show this, we only need to check the $vv$-component of equation  (\ref{TEMeq}) and consider the coefficients of the terms of different orders in $1/ r$. This works because each term of the equation is of the form $F(v)/r^{w}$, with $w\geq 2.$ Hence, to satisfy the equation, each term needs to be equated to zero. 

The $vv$-component of the stress-energy tensor for the Husain type metric reads:
\begin{equation} \label{Tvv_hussain}
    T_{vv}=\frac{\partial_v m}{4\pi r^2}-\frac{\partial_v \psi}{8\pi r^{2k+1}}+\frac{2k-1}{8\pi}\frac{\psi(v)}{r^{2k+2}}f(r,v).
\end{equation}

Taking into account the $vv$-component of (\ref{TEMeq}), 
\begin{equation}\label{vvE}
    \xi^v\p_v T_{vv}+\xi^r\p_rT_{vv}+2T_{vv}=0.
\end{equation}
The $1/r^2$ order term of this directly produces
\begin{equation}\label{mass-lin}
    \xi^v \partial_v^2 m=0\implies m(v)= r_m v,
\end{equation}
where $r_m$ is the rate of mass accretion, and we have discarded a possible constant that can be added to the mass profile. This type of solution for the mass function (\ref{mass-lin}) is unique for all metrics given by Eqs. (\ref{sph-metric}, \ref{HVMetric}). The neutral Vaidya case is a special case of this with $\psi=0$. To get the form of $\psi(v)$ in a general case, we next focus on the terms of order $r^{-(2k+1)}$ in the equation (\ref{vvE}), and get
\begin{equation}
   \partial_v^2 \psi=\frac{2k-1}{v}\partial_v\psi \, .
\end{equation}
This equation has a simple solution 
\begin{equation}\label{qsoln}
    \psi(v)=Q\frac{v^{2k}}{2k},
\end{equation}
where again we have removed an integration constant by imposing a boundary condition on $\psi$, namely $\psi(v=0)=0.$ It is easy to identify that for the charged Vaidya case, $k=1;\alpha=q$, and  Eq. (\ref{qsoln}) implies $q(v)=Q_r v;$ whence $Q_r$ is a constant. Therefore, we have established that for Husain-type metrics, the existence of HKV implies that the metric must have only linear-in-$v$ mass or charge-like functions.

\section{HKV in Rotating Vaidya spacetimes} \label{sec_Kerr_Vaidya}
We turn our attention to rotating versions of Vaidya solutions now. The rotating Vaidya metrics provide useful insights into astrophysical scenarios.  We consider a simple generalisation of non-rotating Vaidya spacetimes, the Kerr-Vaidya black holes \cite{PhysRevD.7.3590}. However, there exist dynamical spacetimes that are axisymmetric other than the Kerr-Vaidya spacetime \cite{Senovilla_1987}, but we leave the study of such cases for the future.  

\subsection{HKV in Kerr-Vaidya metric}
We consider Kerr-Vaidya spacetime in four spacetime dimensions with both the mass and the rotation parameter being dynamical \footnote{Kerr-Vaidya metrics do not satisfy the classical energy conditions. However this feature may not be considered as a pathology, as the violation of energy conditions might be necessary to consistently probe an object in the trapped region by a distant observer \cite{PhysRevD.102.124032} }. The relevant dynamical spacetime metric is
\begin{equation} \label{FKV-metric}
\begin{split}
  ds^2 &= 2\,dv \, dr-\left[1-\frac{2 m(v) r}{\rho(v,r,\theta)^2}\right] dv^2 - 2a(v) sin^2\theta \, dr \,d\phi \\&- \frac{4 m(v) a(v) r}{\rho(v,r,\theta)^2}  \sin^2\theta \, dv \, d\phi  + \rho(v,r,\theta)^2 \, d\theta^2  \\ &+\frac{[\left(r^2+a(v)^2\right)^2-\Delta(v,r) a(v)^2 sin^2\theta]}{\rho(v,r,\theta)^2} \,sin^2\theta \, d\phi^2 \, ,
  \end{split}
\end{equation}
where,
\begin{align*}
  \rho(v,r,\theta)^2 &= r^2 +a(v)^2\, cos^2\theta , \\ \Delta(v,r) &= r^2 - 2m(v)r+a(v)^2.
\end{align*}

Interestingly, this spacetime also admits the same solution for CKE.  
To determine the CKVs, we start with the following ansatz for $\xi$
\begin{equation}
    \xi^a =\{\xi^v(v,r),\xi^r(v,r),0,\xi^\phi(v,r)\} \, ,
\end{equation}
where we have turned on the $\phi$-component of CKV, keeping in mind the Killing vector field for a stationary Kerr BH spacetime. Next, we put this ansatz into the CKE (\ref{CKE3a}), and obtain equations for the components of the CKV, the mass function $m(v)$, and the rotation function $a(v)$. We will aim to solve them using arguments similar to those for the Husain spacetime, worked out in \S\ref{HKV_Husain_Kinematics}. 

The $rr$-component of \eqref{CKE3a} reads 
\begin{equation}
\partial_r\xi^v(v,r)-a(v) \sin ^2(\theta ) \partial_r \xi^\phi(v,r) =0 \, ,
\end{equation}
where each of the two terms on the LHS has to vanish individually, giving us 
\begin{equation}
    \xi^v(v,r)= \xi^v(v)\, , ~~ \text{and}~~ \xi^\phi(v,r)= \xi^\phi(v) \, .
    \end{equation}
Using them back in the equations and following a similar strategy, i.e., the coefficients of $\cos^2\theta$ or $\sin^2\theta$ should vanish independently, we obtain 
\begin{equation}
\begin{split}
 a(v) \left(\partial_r \xi^r(v,r)+\partial_v \xi^v(v)\right)-2 \, \partial_v \xi^v(v) \,  \partial_v a(v)&=0 \, ,\\
 r^2 \left(-\partial_r \xi^r(v,r)-\partial_v \xi^v(v)\right)+2 \, r \,  \xi^r(v,r) &=0 \, ,
\end{split}
\end{equation}
from the $\theta\theta$-component of \eqref{CKE3a}, which will get us 
\begin{equation}
\begin{split}
 \xi^r(v,r) &= r \, \partial_v \xi^v(v) \, , ~~
 { \partial_v \xi^v(v) \over \xi^v(v) } ={\partial_v a(v) \over  a(v)} \, .
 \end{split}
\end{equation}
After feeding these relations back in \eqref{CKE3a}, and following a similar strategy as mentioned above, the $\phi\phi$-component of \eqref{CKE3a} leads us to 
\begin{equation}
\begin{split}
 { \partial_v m(v) \over m(v) } &={\partial_v a(v) \over  a(v)} \, ,
 \end{split}
\end{equation}
which, upon further use in $vv$ and $vr$-components of \eqref{CKE3a}, yields
\begin{equation}
    \partial_v^2 \xi^v(v) =0 \, , ~~\text{and} ~~ \partial_v \xi^\phi(v) = 0 \, .
\end{equation}
From these equations, we immediately obtain that $\xi^v(v)$ is a linear function of $v$, and $\xi^\phi(v) = c_1$, which is an undetermined constant. Therefore, we can summarize the conditions as follows 
\begin{equation}
\begin{split} \label{KerrVaidya_cond}
    \partial_v^2 \xi^v(v) =0 \, ,~~ \partial_v\xi^\phi(v) = 0 \, , ~~\xi^r(v,r) &= r \, \partial_v \xi^v(v)\, , \\
    { \partial_v m(v) \over m(v) } ={\partial_v a(v) \over  a(v)}={ \partial_v \xi^v(v) \over \xi^v(v) } \, .
    \end{split}
\end{equation}
In \eqref{KerrVaidya_cond}, the final condition should be compared with the similar condition that we obtained previously for the Husain spacetime in \eqref{condition_CKV}. In fact they are structurally identical. A similar structure was found by the authors of \cite{Koh:2020hta} in a charged Vaidya BH.\\ 

It is straightforward to argue from \eqref{KerrVaidya_cond} that both the mass function and the rotation parameter function will be a linear function of $v$, i.e., 
 \begin{equation}  \label{ma_function}
    m(v) = M +\mu (v-v_0) \, , ~~     a(v)=a_0+a_1 (v-v_0) \, ,
 \end{equation}
along with the following condition 
  \begin{equation} \label{M_mu_relation}
     \frac{M}{\mu}=\frac{a_0}{a_1} \, .
 \end{equation}
The condition \eqref{M_mu_relation} is actually a consequence of \eqref{KerrVaidya_cond}. Particularly, it is the condition at $v=v_0$ for the parameters of the theory emerging from \eqref{KerrVaidya_cond}. Note that the functions in \eqref{ma_function} imply a situation where we have a stationary Kerr BH initially for $v<v_0$, with mass $M$, and rotation parameter $a_0$, where both are constants. The BH becomes dynamical for $v>v_0$, where the rate of change of mass and the rotation parameters are given by $\mu$ and $a_1$ respectively. It is important to note that the existence of a CKV requires the condition in \eqref{M_mu_relation} to be satisfied.

Next, choosing the integration constants properly, subject to relevant boundary conditions, we find the forms of CKV as
 \begin{equation} \label{CKV_expr1}
     \xi^v(v)=\frac{m(v)}{M} \, , ~~ \xi^r(r)=\frac{\mu r}{M} \, ,~~ \text{and}~~ \xi^\phi(v)= \text{constant}. 
 \end{equation}
The constants have been chosen such that at the static limit, $\xi$ reduces to a Killing vector field of Kerr spacetime with unit norm at asymptotic infinity. Some remarks on the boundary conditions are in order. From \eqref{KerrVaidya_cond} we can see that $\xi^v(v)$, $m(v)$, and $a(v)$ are proportional to each other
 \begin{eqnarray}
     \label{eqn1}
      \xi^v(v)&=&c^\prime_3\, a(v), ~ a(v)=c^\prime_4 \, m(v)\nonumber \\
 ~~&\Rightarrow& ~~ \xi^v(v)= c^\prime_3\, c^\prime_4 \,  m(v)=c^\prime \, m(v).
 \end{eqnarray} 
In writing the functions in \eqref{ma_function}, we have assumed the boundary condition that (i) at $v\rightarrow v_0$, or at $\mu \rightarrow 0$, and $a_1 \rightarrow 0$, they reduce to those of a stationary Kerr BH metric, such that $m(v)=M, ~a(v)=a_0$, both being constants.
 Also, the condition in (\ref{M_mu_relation}), i.e., $M/\mu = a_0/a_1$, is a result of (\ref{KerrVaidya_cond}), and (\ref{M_mu_relation}). This, together with equation \eqref{eqn1} suggests that 
\begin{equation}
    c^\prime_4=\frac{a_0}{M}.
\end{equation}
Next, we make a choice of the constant $c_3$ such that
\begin{equation}
c^\prime= {c^\prime_3}{c^\prime_4}=\frac{1}{M},~~ \Rightarrow{} ~~   \xi^v(v)=\frac{m(v)}{M} \, ,
\end{equation}
as in (\ref{CKV_expr1}) The choice here is directed by the second boundary condition that we used, which is \textit{(ii) to normalize the CKV to unity at asymptotic infinity in the static limit $\mu \rightarrow 0$}. The rest of the relations in  (\ref{CKV_expr1})  follow directly from (\ref{KerrVaidya_cond}) once this $\xi^v$ is obtained. 

Now, if we compute the RHS of \eqref{CKE2} with this CKV, then we find
\begin{equation}
    (\nabla_c \xi^c) =2 \frac{\mu}{M}=2 \frac{a_1}{a_0}~ 
   \Rightarrow ~  \mathcal{L_\xi} g_{ab}= \frac{\mu}{M} g_{ab}=\frac{a_1}{a_0} g_{ab}. 
\end{equation}
Therefore, we confirm that $\xi$ is a homothetic Killing vector.

One can also prove that homotheticity implies the same HKV as those obtained for other spacetimes. For Kerr-Vaidya BH the condition (\ref{div}) leads to the following equation
\begin{equation}\label{Kerr-div}
    \frac{2aa_v\cos^2\theta \xi^v}{r^2+a^2 \cos^2 \theta}+ \frac{2r\xi^r}{r^2+a^2 \cos^2 \theta}+\p_v\xi^v+\p_r\xi^r=4c \, .
\end{equation}

Next, the $vr$-component of equation \eqref{TEMeq} reads $\xi^v=\tau(v)$. Further, equation  (\ref{thetaeq}) leads,
\begin{equation}\label{Kerrtheta}
    2\xi^vaa_v\cos^2\theta+2 \xi^r r=2c(r^2+a^2 \cos^2 \theta).
\end{equation}

Comparing both sides of the above equation (\ref{Kerrtheta}) forces $\xi^r=cr.$ Further,  replacing the left hand side of equation  (\ref{Kerrtheta}) into (\ref{Kerr-div}) we get
\begin{equation}
    3c+\p_v\tau(v)=4c \implies \xi^v=cv.
\end{equation}
 
In deriving this, we only assumed that we are working with the Kerr-Vaidya metric in \eqref{FKV-metric}. Thereafter, demanding that we have an HKV that satisfies \eqref{div}, fixes $\xi^a=\{c\,v,\, c\, r,\, ,0,\, 0\}$, up to additive constant factors. Moreover, we have also been able to write down the expression for a stress-energy tensor $T_{ab}$ for Kerr-Vaidya space-time satisfying Einstein's equations. We have also been able to show that $T_{ab}$ satisfies \eqref{TEMeq} which is a consequence of having an HKV $\xi^a=\{c\,v,\, c\, r,\, ,0,\, 0\}$, given that the $m(v)$ and $a(v)$ are also linear functions of $v$, as given in \eqref{ma_function} with the condition \eqref{M_mu_relation}, paralleling the analysis that we performed for the Husain type metrics in \S \ref{HKVform}. We performed this analysis in Mathematica, and the analytical expressions for $T_{ab}$ are not illuminating enough in this case, unlike \eqref{Tab_hussain} or \eqref{Tvv_hussain}. Hence, we do not write them here to avoid the clutter.

Therefore, to summarize, it can be argued that the linear dependence of mass and angular momentum in $v$ follows directly from the existence of HKV for a Kerr-Vaidya space-time. In other words, most importantly, we also learn that for a Kerr-Vaidya BH, which only has a dynamical mass function but the angular momentum is non-dynamical, i.e., a constant, one can not have a CKV. It is crucial for the existence of CKV that we have both dynamical mass and angular momentum, written in \eqref{ma_function}, along with the condition in \eqref{M_mu_relation}. It is intriguing that Eq. \eqref{TEMeq}, with the above HKVs, is solved using the same types of mass and rotation parameters as in \eqref{ma_function}, and the parameters must again satisfy the condition \eqref{M_mu_relation}

This dynamical Kerr-Vaidya spacetime ($g_{ab}$) (as well as the Husain type metric) can be mapped to a conformally stationary (static) spacetime $\left(\Omega^{-2}g_{ab}\right)$ obeying equation \eqref{cnftr}. The HKVs will be the Killing vectors of that stationary spacetime satisfying the Killing equation (\ref{KE}). As we stated earlier, the value of the conformal factor $\Omega$ is not unique. We have found, from Eq. \eqref{conf-factor}, the following choices agree with the required conditions.
\begin{equation}\label{cf3} 
  \Omega^2_{(1)} = \frac{m(v) r}{2 M^2} \, , ~~
  \Omega^2_{(2)} = \frac{m(v)^2}{M^2} \, ,~~
   \Omega^2_{(3)} = \frac{2 m(v) \mu r}{M^2} \, . 
\end{equation}

Therefore, using the conformal factor, one can change the dynamical metrics to a static/stationary spacetime, as in \eqref{conf_dynamic_to_static}, i.e.,
\begin{equation}
    \label{dynamic-stat}
    ds^2_{dynamic} = \Omega^2 ds^2_{static/statinary}.
\end{equation}
It must be mentioned that the conformally related static or stationary spacetimes are not solutions of the same Einstein's equation that the original dynamical metrics satisfy; they are the solutions of conformally transformed Einstein's equation.

\subsection{HKH in a slowly rotating Vaidya black hole}
In this section we turn our attention to a slowly rotating Kerr-Vaidya metric to display some properties of HKH explicitly due to a better analytic control. A slowly rotating Kerr BH provides a useful model to study many astrophysical phenomena, including thin shell collapse. The metric is obtained by neglecting the $a^2$ and higher-order terms in a Kerr-Vaidya metric. We consider the slowly rotating Vaidya metric, with both mass and rotation parameters as functions of the advanced null coordinate $v$, 

\begin{equation}\label{sl-Kerr}
\begin{split}
    ds^2=&-\left(1-\frac{2m(v)}{r}\right)dv^2+2 dv dr-2 a(v) \sin^2 \theta dr d\phi \\  -&\frac{4m(v) a(v)}{r}\sin^2 \theta dv d\phi+r^2(d\theta^2+\sin^2\theta~ d\phi^2) \, .
\end{split}
\end{equation}

The matter stress tensor can be worked out for the above metric (\ref{sl-Kerr}) with the field equation
  \begin{equation}
      R_{a b}-\frac{1}{2} g_{ab} R = 8 \pi  T_{ab} \, ,  ~~~~(G=1) \, .
  \end{equation}
  In the slow rotation limit, there exists a vector, 
  \begin{equation}
    v^a =\left(1,0,0, \frac{a(v)}{2 r^2}  \right) \, ,
  \end{equation}
  such that $v^a v_a = 0 + \mathcal{O}(a)^2$ and the stress energy tensor can be written as, 
  \begin{equation}
      T_{ab}= \frac{m'(v)}{4 \pi r^2} v_a v_b+ \mathcal{O}(a)^2 \, ,
  \end{equation}
  up to linear order in $a(v)$ and also imposing $a'(v)=0$ \cite{PhysRevD.1.3220}. Therefore, we can say that the stress energy tensor of Kerr-Vaidya spacetime satisfies some null energy condition up to linear order in the rotation parameter.

The slowly rotating Vaidya metric also has the same HKVs $\xi^a=\{cv+c_0,cr,0,\chi\}$. The proof of this for the metric (\ref{sl-Kerr}) remains the same as shown in (\ref{HKVform}). This follows from the fact that the slowly rotating Vaidya metric bears many similarities with the non-rotating metrics. As the spherical part of this spacetime remains the same, equations (\ref{thetaeq}) and (\ref{div}) will be the same. Further, the mass and rotation parameters both depend on the $v$ linearly. This can also be seen by retracing the same path as depicted from equation  (\ref{vvE}). In fact, since in this limit, the stress tensor becomes identical to the non-rotating Vaidya case, everything follows almost identically to the case for the Vaidya BH. 

\begin{equation}
    T_{vv}^{srV}=\frac{m_v}{4\pi r^2}+\mathcal{O}(a^2).
\end{equation}

One interesting fact for this slow-rotation case is that if one only turns on a dynamical mass with constant rotation, we do not get any solution to the CKE. 

For the mass function and the rotation parameter function as in \eqref{ma_function}, the slowly rotating Kerr-Vaidya spacetime admits a CKV $\xi$, same as the full rotation case, i.e., \eqref{CKV_expr1}, satisfying the CKE \eqref{CKE2}.
 
Additionally, we observed an important fact that if we choose
 \begin{equation} \label{CKV_KV_2}
 \begin{split}
      \xi^v &= \frac{m(v)}{M}, \,\,\,\,\,\,
    \xi^r = \frac{\mu r}{M}, \\
    \text{and} \,\,\,\,\,\,
 \xi^\phi &= \frac{a(v)}{4 M^2} =\frac{a_0 +a_1(v-v_0)}{4M^2} \, ,
 \end{split}
 \end{equation}
i.e., if $\xi^\phi$ is set as the angular velocity of the BH up to linear order in the dynamical rotation parameter $a(v)$ (an analogy with the Killing vector field of the stationary Kerr spacetime metric), then also the CKE \eqref{CKE2} is satisfied up to linear order in $a_1$. Importantly, here also, the same condition in \eqref{M_mu_relation} has to be satisfied.
 
The CKV, written in \eqref{CKV_expr1}, with the choice of $\xi^\phi=0$, will be null at the CKH given by
 \begin{equation}
     \xi^a \xi_a|_{CKH} =0 \, . \label{CKH}
 \end{equation}
 This equation has solutions at 
 \begin{equation} \label{CKH1}
 \begin{split}
   &r|_{CKH}=   \frac{a(v)}{4a_1} \left(1\pm\sqrt{1-16 \mu }\right) \, ,
 \end{split}
 \end{equation}
 which will be real given $ \mu \leq 1/16$. Equation (\ref{CKH1}) denotes the location of the CKH. Also, for the choice of CKV in equation (\ref{CKV_KV_2}), the position of the CKH is the same as given in equation (\ref{CKH1}).

\section{Thermodynamics for Kerr-Vaidya black holes} \label{sec_thermo}
\subsection{Surface Gravity for Rotating Vaidya Black Hole}
Black hole thermodynamics asserts that BHs emit thermal radiation through the production of quantum particles, and this radiation is identified by the Hawking temperature ($T_H$). The surface gravity($\kappa$) of a stationary BH is directly  proportional to its Hawking temperature,
\begin{equation}
    T_H=\frac{\kappa}{2 \pi}.
\end{equation}
In section \ref{IIB}, we introduced the idea and definitions of different kinds of surface gravity on a conformal Killing horizon. After computing the three $\kappa$'s for slowly rotating Kerr-Vaidya spacetime (\ref{sl-Kerr}), we found that they satisfy equations (\ref{kappa11relation}) and (\ref{kappa123relation}), up to linear order in $a_1$ for $\xi^a = \left(\frac{m(v)}{M},\frac{\mu r}{M},0,\frac{a(v)}{4 M^2}\right)$.

If we take the linear mass function and the rotation parameter function as in equation \eqref{ma_function}, subjected to the boundary condition (\ref{M_mu_relation}), then the calculation of $\kappa_1$, $\kappa_2$ and $\kappa_3$ at CKH (\ref{CKH1}) turns out to be the same for both of the above mentioned cases. The values are given by
\begin{equation}\label{kappa123}
\begin{split}
   \kappa_1 &= \frac{a_1 \left(-16 \mu +\sqrt{1-16 \mu }+1\right)}{8 a_0\mu } \, ,\\
    \kappa_2 &=\frac{{a_1} \left(\sqrt{1-16 \mu }+1\right)}{8{a_0} \mu } \, ,\\
    \text{and} ~~ \kappa_3 &= \frac{16 {a_1} \mu }{{a_0} \left(\sqrt{1-16 \mu }-1\right)^2} \, .
    \end{split}
\end{equation}
Also, we obtained 
\begin{equation}
    \lambda= \frac{1}{4} ({\nabla}_b \xi^b)= \frac{a_1}{a_0} \, ,
\end{equation}
where $\lambda$ is defined in \eqref{CKE3}. From here, a few steps of calculation yields
\begin{equation} 
    {\kappa}_1 = {\kappa}_2 - 2\lambda = {\kappa}_3 -\lambda \, .
\end{equation}
As we have explained before in section \ref{IIB}, $\kappa_1$ manifests the conformally invariant surface gravity for slowly rotating Kerr-Vaidya spacetime (\ref{sl-Kerr}) (it will be the same when calculated at the KH of the conformally mapped stationary Kerr-Vaidya spacetime)  and thus also gives an interpretation of conformally invariant temperature of the dynamical BH.  The norm of the relevant CKVs, either (\ref{CKV_expr1}) or (\ref{CKV_KV_2}), blows up when computed at asymptotic infinity. Therefore, surface gravity should be measured by an observer following the trajectory of CKVs, unlike at $r\rightarrow \infty$ in the stationary case. For that, the CKVs have to coincide with the normalized four velocity of the observer \cite{Nielsen:2017hxt}.

The existence of a CKV in Vaidya-like dynamical metrics enables one to define quantities that mimic the thermodynamic variables of a BH. As already presented, there can be three definitions of geometric surface gravity of a space-time Eq. \eqref{kappa123}. All of them are equivalent on a KH. However, they do not match on the CKHs. Not all of these definitions are accepted as a proper definition of the surface gravity of BHs. Sometimes, $\kappa_2$ is reckoned as the desired definition of surface gravity, which is inspired by the operational definition of surface gravity, i.e., related to the force needed to be exerted on a unit test particle at the horizon. However, the first of these, i.e., $\kappa_1$, remains unaltered whether we use a Killing vector or a Conformal Killing vector of a spacetime, and thus this one is usually referred to as the conformally invariant temperature. The cases described in this note are all HKHs, and in these cases, we get some additional advantages to define thermodynamic quantities. First of all, for HKV, the surface gravities differ from each other by a constant factor. Therefore, the surface gravity scales up or down along Homothetic Killing trajectories by a constant factor. Secondly, the zeroth law for the generic Vaidya spacetimes we have considered, including Kerr-Vaidya spacetimes, reads 
\begin{equation}
    \mathcal{L}_\xi[\kappa_2- 2\lambda] = 0.
\end{equation}
In addition, the surface gravity $\kappa_1$ is constant. This is because of the fact the CKV is actually a Homothetic Killing vector.

\subsection{Temperature and First Law}

In stationary spacetimes, the surface gravity is usually identified as the thermodynamic temperature up to a factor. For dynamical spacetimes, we do not have such a simple mapping between the geometric temperature or the surface gravity and the thermodynamic temperature \cite{Nielsen:2007ac, Nielsen:2017hxt, Hammad:2018ddv}. One needs to calculate the Hawking temperature by any acceptable method available in the literature to compare these two. Since a Killing vector that defines the surface gravity or temperature for a Killing horizon must also satisfy the desired normalized time-like behavior at the far region, one may rescale the Conformal (Homothetic) Killing vector accordingly. So we are looking for one $\xi'^{a}=c\,\xi^{a}$, for which $\xi'^{a}\xi'_a\to -1$ as we approach the static or stationary limit. For VRN spacetime, this means
\begin{equation}
\begin{split}
    &\xi'^{a}
    =\frac{1}{\sqrt{-\xi^{a}{\xi}_{a}}}\xi^{a}\\
    &=\frac{Mr}{\sqrt{m(v)^2r^2-2 m(v)^3r+2m(v)^2 q(v)^2-2\mu m(v)r^3}}\xi^{a} \, ,
\end{split}
\end{equation}
is the normalized HKV that will generate a surface gravity that would be dependent on the position. In the conformally static spacetime, the surface gravity $\tilde{\kappa}'_1$, w.r.t. the normalized HKV $\xi'^{a}$ would acquire a conformal factor accordingly \cite{Nielsen:2017hxt}.
\begin{equation}
    \tilde{\kappa}'_1=c \, \Omega \, \tilde{\kappa}_1 =c \, \Omega \, \kappa_1\, ,
\end{equation}
where $\tilde{\kappa}_1$ denotes the surface gravity of the metric $\Omega^{-2} g_{ab}$, with $g_{ab}$ being the VRN metric as in \eqref{metric_VRN}. 
This surface gravity is measured by an observer traveling along the flow generated by the HKV. Similar expressions should also hold for rotating Vaidya-like spacetimes. 

There are many approaches to defining a thermodynamic temperature in such dynamical spacetimes.
In \cite{Nielsen:2017hxt}, an \textit{empirical} definition of temperature has been proposed at the HKH, inspired by the Bekenstein-Hawking entropy relation, 
\begin{equation}
   T_{eff}= \frac{\delta m(v)}{\delta A/4}=\frac{4\mathcal{L}_\xi m(v)}{\mathcal{L}_\xi A} \, .
\end{equation}

This definition resembles an analogous definition of flux temperature $T_F$ in non-equilibrium thermodynamics, defined as the ratio of flux of energy and entropy that  passes through a surface enclosed by a volume $V$ containing some radiation \cite{landsberg1980thermodynamic, casas2003temperature},  
\begin{equation}
T_F=\frac{\dot{E}}{\dot{S}}
,
\end{equation} where a dot denotes the rate of change of energy or entropy. In a strict equilibrium situation, the fluxes are zero. This means the BH is inert and not even allowing a small perturbation from the stationarity. For the non-stationary case, the flux temperature should reach the value of the Hawking temperature in the appropriate limit, for example $\simeq 1/(4M)$, as expected for the Schwarzschild-Vaidya case. In the dynamic phase, the temperature is proportional to $v^{-1}$, showing its dynamic character. However, this definition lacks a proper justification in terms of some covariant quantity defined on the CKH for more general Vaidya-type spacetimes. In general, for Vaidya spacetimes admitting spherical symmetry, a quasi-local energy can be defined with respect to the Kodama vector $K^a=-\epsilon^{ab}\nabla_b r$ \cite{10.1143/PTP.63.1217}. The conserved charge corresponding to the Kodama vector on a trapping horizon is known as the Misner-Sharp-Hernandez (MSH) energy \cite{Misner:1964je, PhysRevD.53.1938, Abreu:2010ru}, and is given by:

\begin{equation}
    E(v,r)=\frac{r}{2}(1-\nabla^ar\nabla_ar).
\end{equation}

For the Vaidya solution, $E(v,r)=m(v)$. For VRN this becomes \[E(v,r)=m(v)-\frac{q(v)^2}{2r}.\] The change of this energy along the apparent horizon serves to produce the generalized first law for dynamical spacetimes \cite{PhysRevD.53.1938}.  This MSH energy serves the purpose of establishing a flux law that mimics the first law of thermodynamics \footnote{Kerr-Vaidya black hole with constant rotation parameter $a$ also supports a Kodama vector, but we don't have any evidence of possessing a Kodama-like vector when $a$ is a function of $v$ \cite{Dorau:2024zyi}.}. One can check that at HKH
\begin{equation}\label{fluxlaw}
 T_{eff}= \frac{4\mathcal{L}_\xi E(v,r)}{\mathcal{L}_\xi A}\Big{|}_{HKH}=\frac{4E(v,r_{HKH*})}{\mathcal{L}_\xi A}\simeq{1\over v}F(\mu,\lambda),
\end{equation}
where $F(\mu,\lambda)$ is a function of the accretion rate parameters for mass and charge, and they have a form that reduces to the desired Hawking temperatures in the static limit. Here, $r_{HKH*}$ is the HKH that has a finite static limit. This can be explicitly checked for the Husain-type metrics. This relation \eqref{fluxlaw} can be regarded as a first law for HKH in a spherical setting, $\delta E(v,r)=T_{eff}~ \frac{\delta A}{4}$.

It is easy to see by direct calculation of expansions for the radially inward and outward geodesics $l^a$ and $n^a$ that the HKH or CKH are not trapped surfaces as the Killing horizons used to be. One of the obstacles to establishing the laws of thermodynamics on HKH is this untrapped nature. Under a quasi-local setup, there is some success in this direction. In this setup, assuming that the CKV generates a shear-free null congruence, one can prove a first law for CKH \cite{Chatterjee:2014jda, Chatterjee:2015fsa}. However, there are no known solutions for which this has been tested so far.  We may focus on the thermodynamics of the conformally stationary or static spacetimes, but those spacetimes are solutions of the conformally transformed Einstein equation. Further, these spacetimes are not in general asymptotically flat, as can be seen from Eqs. \eqref{conf-trans}, \eqref{dynamic-stat}. So, extra care needs to be taken to address their thermodynamic properties. For example, the $r$ coordinate fails to be identified as an areal radius in these metrics. Further, the static or stationary metrics that are obtained with the help of a conformal transformation don't automatically capture the desired features of the original dynamical spacetimes. One should note that these conclusions are only valid within the restricted region between the HKHs, and the question of Hawking temperature can't be addressed properly.

As the different notions of temperatures for Vaidya spacetimes are available from geometric considerations, it is not clear which of them can be regarded as the thermodynamic temperature related to a thermal particle spectrum. To this end, we can resort to the study of particle creation by a self-similar Vaidya-type spacetime. This might provide us with some insight into the Hawking radiation and its temperature. The first step to carry out such an analysis is to locate clearly the future and past null infinities for dynamical spacetimes, as the asymptotic state needs to be defined there. In general, this is not a trivial task, as null Cauchy surfaces and singular surfaces crop up in the maximal analytic extended spacetime \cite{Hiscock:1982pa}. In what follows, we show the construction of a conformal completion of a fully charged Vaidya spacetime that can unambiguously attach future and past null infinities.
\vskip -2cm

\subsection{Maximal analytic extended charged self-similar Vaidya solution}

In the following, we aim to explore another interesting aspect of the dynamical horizon, namely, to study the creation of massless scalar particles in spacetimes that contain shell-focusing singularities using the setup of CKH for Vaidya BH developed so far. Vaidya solutions are known to be of a self-similar nature, i.e., the metric gets scaled by an overall factor under the simultaneous scaling $v \to \lambda\,v, \, r\to \lambda \,r$, with $\lambda$ being a constant parameter \cite{Nolan:2006pz}. Such solutions can produce naked or marginally naked singularities \cite{Hiscock:1982pa}. As studied in  \cite{Hiscock:1982pa}, a maximally extended Vaidya spacetime will have a Cauchy horizon at the location of the HKH, and it also develops singularities. To resolve this problem, a spherically symmetric collapsing matter was considered so that a curvature singularity remains just inside the entire event horizon, with no effect outside it. In \cite{Hiscock:1982pa}, the issue was addressed using a Schwarzschild-Vaidya metric. We want to explore this description for Reissner–Nordström-Vaidya spacetime. \\

 Let us start by writing again the metric for the spherically symmetric charged Vaidya spacetime,
\begin{align}
    d{s}^2 = &-\left[1-\frac{2 m(v)}{r} + \frac{q^2(v)}{r^2}\right] dv^2 + 2~dv~dr\\ \nonumber &+ r^2 \left(d\theta^2 + \sin^2 \theta ~d\phi^2 \right)  \, ,
 \end{align}
 with the mass and charge functions given by
 \begin{equation} \label{mass_condition}
 m(v)=
     \begin{cases}
          0 & \text{if } v \leq 0  \, ,\\
    \mu \, v  & \text{if } 0 \leq v \leq v_0 \, ,\\
    M & \text{if } v \geq v_0 \, ,
     \end{cases}
 \end{equation}
 and 
  \begin{equation}
 q(v)=
     \begin{cases}
          0 & \text{if } v \leq 0 \, ,\\
    q_1 \,  v  & \text{if } 0 \leq v \leq v_0 \, ,\\
    Q & \text{if } v \geq v_0 \, .
     \end{cases}
 \end{equation}
 As we see from above, $m(v)$ and $q(v)$ are linear functions of $v$  in $0 \leq v \leq v_0$. In this window, the HKV 
 \begin{equation} \label{ckv}
     \xi^a =(v,r,0,0) \, ,
 \end{equation} 
 satisfies the CKE
 \begin{equation} \label{cke}
     \mathcal{L_\xi} g_{ab}= \nabla_a \xi_b +\nabla_b \xi_a = 2g_{ab} \, .
 \end{equation}
 Next, we construct a double null coordinate system in the three segments of the model spacetime:\\
 
\noindent\textbf{ I. Minkowskinan region ($v\leq 0$):}\\
 We define a new coordinate 
 \begin{equation}
     U= v-2r
 \end{equation}
 in terms of which the spacetime metric is given by 
 \begin{equation}
     ds^2= -dU~dv + r^2 \left(d\theta^2 + \sin^2 \theta ~d\phi^2 \right).
 \end{equation}
\textbf{II. Null fluid region ($0 \leq v \leq v_0$):}\\
 Here, we first introduce the new coordinates
\begin{equation}
     z=\frac{v}{r} \, , ~~ \text{and}~~  \zeta=ln~v \, .
 \end{equation}
 In terms of these coordinates, the metric in the second segment takes the following form 
 \begin{align}
     ds^2= e^{2\zeta}\left[-\left(1-\frac{2}{z}-2 \mu z+ q_1^2 z^2\right)d\zeta^2 \right. \nonumber \\  \left. \quad -\frac{2}{z^2}d\zeta dz +\frac{1}{z^2}d\Omega_2^2\right] \, .
 \end{align}
 Next, we perform another coordinate transformation to introduce $\eta$ defined as 
 \begin{equation}
     \eta=\zeta-2 z^{*} \, ,
 \end{equation}
 such that
 \begin{equation} \label{dzstar}
     dz^{*}= \frac{dz}{-z^2\left(1-\frac{2}{z}-2 \mu z+ q_1^2 z^2\right)} \, .
 \end{equation}
 Finally, the spacetime metric can be written as 
 \begin{equation}
  ds^2= e^{2\zeta}\left[-\left(1-\frac{2}{z}-2 \mu z+ q_1^2 z^2\right)~d\zeta~d\eta +\frac{1}{z^2}d\Omega_2^2\right] \, ,
 \end{equation}
 which is the desired double null coordinate system where $z$ is an implicit function of $\zeta$ and $\eta$. The null surfaces, where the norm of CKV (\ref{ckv}) vanishes, will be defined by the vanishing of $g_{\eta \zeta}$, that is,
 \begin{equation}
    1-\frac{2}{z}-2 \mu z+ q_1^2 z^2=0 \, . 
 \end{equation}
This is a cubic equation of $z$, which admits three solutions, which are real if we impose the condition $\mu \leq 1/16$. If we consider the charge accretion rate $q_1$ to be very small, then the solutions read up to linear order in $q_1$ as
\begin{equation}
\begin{split}
    z_1 &= \frac{2 \mu }{{q_1}^2}-\frac{1}{2 \mu }+\mathcal{O}\left(q_1^2\right) \, , \\
    z_2&= \frac{1}{4 \mu}\left[1+\sqrt{1-16\mu}\right] +\mathcal{O}(q_1^2) \, , \\
  z_3& =  \frac{1}{4 \mu}\left[1-\sqrt{1-16\mu}\right] +\mathcal{O}(q_1^2) \, ,
  \end{split}
\end{equation}
which indicate the locations of the CKH or homothetic killing horizons.

\begin{figure}[h]
\includegraphics[width=8cm]{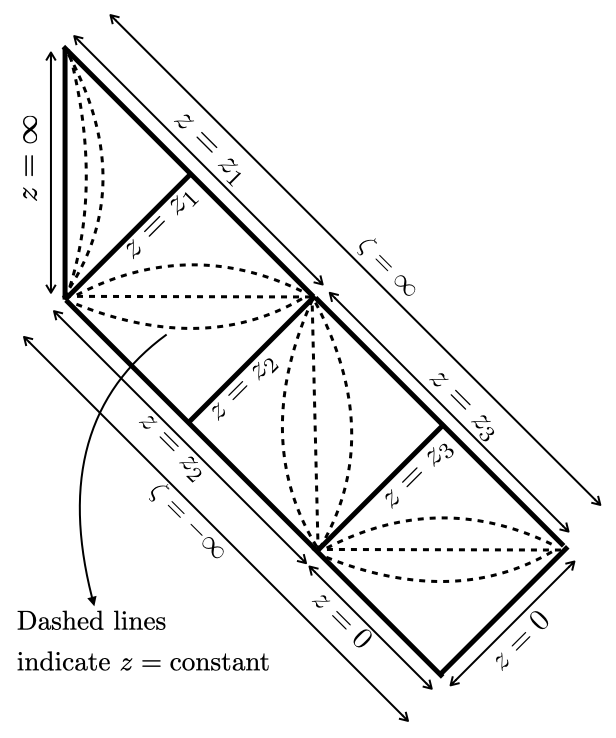}
\caption{Maximum analytic extension of the self-similar (homothetic) Charged Vaidya metric except for the region $m\leq 0$. Dashed lines indicate surfaces of constant $z$.} \label{figure1}
\end{figure}

The Kretschmann scalar can also be computed as follows 
\begin{equation}
\begin{split}
    R^{abcd} R_{abcd} &= \frac{8 \left(-12 r m(v) q(v)^2+6 r^2 m(v)^2+7 q(v)^4\right)}{r^8} \\ &=\frac{48 \mu ^2 v^2}{r^6}-\frac{96 \mu  q_1^2 v^3}{r^7}+\frac{56 q_1^4 v^4}{r^8} \, ,
\end{split}
\end{equation}
which implies the existence of curvature singularities at $\zeta=-\infty$, $z=\infty$. Integrating equation (\ref{dzstar}), the tortoise-like coordinate $z^*$ can be obtained as 
\begin{equation}
\begin{split}
   & z^*= \frac{1}{q_1^2}\left[\frac{\ln z}{z_1z_2z_3}-\frac{\ln \left|z-z_1\right|}{z_1(z_1-z_2)(z_1-z_3)}\right.  \\  &\left. +\frac{\ln \left|z-z_2\right|}{z_2(z_1-z_2)(z_2-z_3)}-\frac{\ln \left|z-z_3\right|}{z_3(z_1-z_3)(z_2-z_3)}\right] \, .
\end{split}
\end{equation}
 Therefore, we find that the singularities occur at $z=0,~ z=z_2$ ($\zeta\rightarrow -\infty$), see Fig.\ref{figure1}.

 The outermost homothetic killing horizon $z_3$ is a Cauchy horizon. Since we want to obtain a model spacetime that is asymptotically flat and vacuum in the distant future, the homothetic charged Vaidya solution should be cut along the null surface $v=v_0$ (constant) ($\zeta=$ constant) and attached to a portion of a Reissner–Nordström spacetime. The mass ($M$) and the charge ($Q$) of the final RN BH will be such that the Cauchy horizon coincides with the outer event horizon $(r_+)$ of the BH, 
\begin{equation}
   z_3= \frac{v_0}{r_+}= \frac{v_0}{M+\sqrt{M^2-Q^2}} \, . 
\end{equation}

This indicates a null shell structure present at $v=v_0.$ Usually, the HKHs are outside the critical surfaces that may trigger the question of energy condition violation. We may restrict ourselves to a region for the mass and charge parameters such that the singularities can be visible only locally; otherwise, there can be global naked singularities for which the event horizon lies inside the Cauchy horizon $z=z_3$ surface. The coincidence of the Cauchy horizon with the event horizon of the RN black hole indicates a marginally naked character of such a spacetime.

\begin{figure}[h]
\includegraphics[width=7cm]{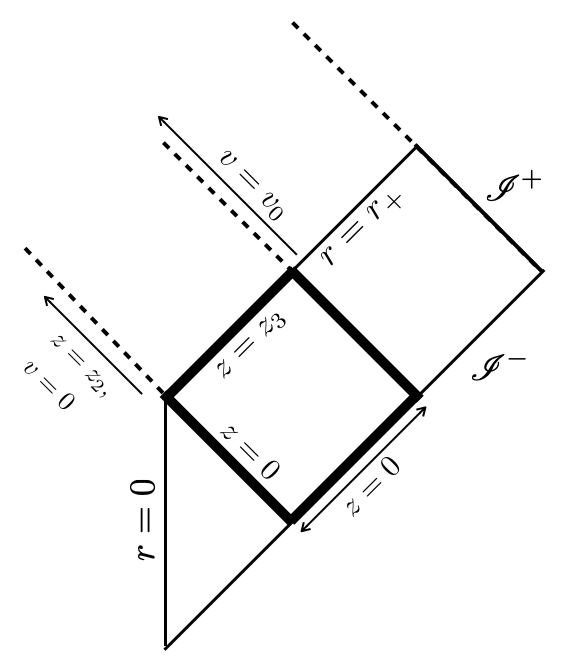}
\caption{The region $v \leq 0$ is described by the Minkowski spacetime metric; the $v\geq v_0$ region is given by the RN spacetime metric with constant mass $M$ and charge $Q$. In the central region $0\leq v \leq v_0$, the homothetic charged Vaidya null fluid has collapsed.} \label{figure2}
\end{figure}
\textbf{III. Reissner–Nordström region ($ v \geq v_0$):} 
 Here, the usual double null coordinates (i.e., the retarded and advanced time coordinates) are considered, 
 \begin{equation}
     u=v-2r^{*},
 \end{equation}
 such that 
 \begin{equation}
   dr^{*}= \frac{dr}{\left[1-\frac{2 M}{r} + \frac{Q^2}{r^2}\right]} \, . 
   \end{equation}
In these coordinates, the metric takes the known form
\begin{align}
 d{s}^2 = -\left[1-\frac{2 M}{r} + \frac{Q^2}{r^2}\right] du~dv + r^2~d\Omega_2^2 \, ,
 \end{align}
where $r$ is an implicit function of $u$ and $v$. 

Existence of a complete $\mathscr{I}^+$ guarantees (see Fig.\ref{figure2}) that one can unambiguously study the Hawking effect by using the standard mode decomposition method. Further, the existence of an HKV in the region $0\leq v\leq v_0$ enables us to separate a massless scalar wave equation, $\Box \Phi=0$, in this background. We do not outline the Hawking spectra for the VRN spacetime here, but we believe the outcome of such an analysis would show a departure from thermality as shown for the Schwarzschild-Vaidya case in \cite{Hiscock:1982pa}. For rotating Vaidya BH, a similar method may be applied; however, it has much less analytic control. Since the RN metric shares many common features of Kerr spacetime, studying the scattering problem for the VRN case using the maximally extended setup might provide some useful insights into the issue of evaporation of BHs for a rotating case. One may still consider the slowly rotating case; the analytically extended metric in the infalling radiation zone now reads
\begin{equation}
    \begin{split}
     ds^2= e^{2\zeta}\bigg[&-\left(1-\frac{2}{z}-2 \mu z\right) ~d\zeta~d\eta \\&  -\left(\frac{2 a_1 sin^2\theta}{z} +4a_1 \mu z sin^2 \theta\right) d\zeta~d\phi \\
     &+\frac{2 a_1 sin^2\theta}{z^2} dz~d\phi +\frac{1}{z^2}d\Omega_2^2\bigg].
 \end{split}
 \end{equation}
 However, the results would be almost identical to those of the Schwarzschild-Vaidya case to leading order if we consider only the equatorial plane. 

 \section{Conclusion} \label{sec_conclusions}
 
In this note, the existence of homothetic Killing vectors for generic Vaidya-like spacetimes is established for a general spherical symmetric setup. Further, we extend the existence of HKV  fields for the Kerr-Vaidya spacetimes. The existence of a solution for the conformal Killing equation requires that the mass, angular velocity, and general charge functions all become dynamical, i.e., depend on the `null time'. It is intriguing that, in all such cases, a HKV exists for a linear mass/charge/angular-velocity function. On the other hand, if one demands an HKV in Vaidya-like spacetimes, then one can prove that the forms of such vectors are identical. Further, using the Einstein equation for the matter part, the HKV yields a solution with linear source terms: masses, charge, or angular velocity. However, there can be proper conformal Killing vectors in such spacetimes, but they are going to be supported by other types of mass functions \cite{Ojako:2019gwc}. We have also noticed that for the BH parameters of the form \eqref{malphav2} or \eqref{ma_function}, the existence of CKV demands conditions like \eqref{condition_CKV}, \eqref{KerrVaidya_cond}, or equivalently \eqref{constraint1}, \eqref{M_mu_relation}. On noticing similarities between these conditions, it seems the CKEs for the generic Vaidya metrics considered here may have some universal character. This universality might have an underlying connection with the self-similarity of the metrics. A deeper understanding of this feature would be worth pursuing.

The advantage of identifying a homothetic Killing vector in a dynamical spacetime is that one can map it to a static or stationary spacetime. This enables us to study certain features of the BHs using the Killing horizons of the conformally related spacetime. Although the practical use of such an advantage has not been widely explored. Since the conformally related spacetime can have a very different character (it may not have the same asymptotic structure), the utilization of such constructions needs to be checked carefully. This might yield an interesting result in the study of quasi-normal frequencies for dynamical BHs \cite{Redondo-Yuste:2023ipg, Capuano:2024qhv}. We would like to address this issue in future.

We further propose a dynamical notion of the first law or flux law by expressing the temperature as the ratio of the change in the Misner-Sharp-Hernandez energy to the change in the horizon area. However, thermodynamic temperature may differ. We would like to explore the Wald entropy and the second law for HKHs in the future. To understand the thermal effects for spacetimes admitting HKHs, we first need to study particle creation by the Hawking process. In this regard, we followed the study by Hiscock et al. on the Schwarzschild-Vaidya BH and obtained the maximally extended self-similar charged Vaidya spacetime \cite{Hiscock:1982pa}. Since one can attach a future (or past) null infinity to the dynamical spacetime and define asymptotic states, this simplifies the study of BH perturbations and Hawking radiation \cite{Hiscock:1982pa}. It is known that these evaporation models produce non-thermal radiation from BHs. Therefore, it would be interesting to relate some of the dynamical temperatures derived from geometric analyses to the temperature emerging from the studies of BH evaporation. \\ 

\textbf{Acknowledgements:} The research of SB is partially supported by the SERB-ANRF through the MATRICS grant, number MTR/2022/000170. NK acknowledges support from a recently concluded MATRICS research grant (MTR/2022/000794) from the Anusandhan National Research Foundation (ANRF), India.

\bibliographystyle{apsrev}
\bibliography{ref}
\end{document}